\documentclass[a4paper]{aa}
\usepackage{psfig,times}
\begin{document}
\title{The stability of   
       protostellar disks with   Hall effect and buoyancy}
\author{V.~Urpin\inst{1,2} \and G.~R\"{u}diger\inst{1}}
\offprints{V.Urpin}
\institute{Astrophysikalisches Institut Potsdam, An der 
           Sternwarte 16, D-14482 Potsdam, Germany \and
           A. F. Ioffe Institute of Physics and Technology,
           194021 St. Petersburg, Russia}

\date{\today}

\abstract{
The stability properties of inviscid protostellar disks are examined taking
into account the Hall effect and buoyancy. Depending on the 
parameters, different types of instabilities can exist in different
regions of disks. In a very low ionized region, the instability
associated with baroclinic effects of buoyancy is likely most 
efficient. The Hall-driven shear instability can lead to 
destabilization of regions with a higher ionization. The 
magnetorotational instability modified by buoyancy can only be 
a destabilizing factor in regions with  strong magnetic field or a relatively high 
conductivity ($\sigma B^2/\rho> a_e \Omega$, with $a_e$ the magnetization parameter of electrons).  
}
\authorrunning{V. Urpin \&  G. R\"udiger}
\titlerunning{The stability of protostellar disks with the Hall effect}

\maketitle

\section{Introduction}

The models of astrophysical disks require sufficiently strong 
turbulence to enhance the efficiency of angular momentum transport. 
At present, there is no commonly accepted view point as to how a 
laminar flow in disks is disrupted and turbulence generated. In 
general, turbulence may be generated due to various hydrodynamic
and hydromagnetic instabilities which can arise in differentially 
rotating non-uniform gaseous disks but the exact origin of turbulence 
is still controversial. 

In accretion disks, the origin of turbulence is oftenly attributed 
to the magnetorotational instability  since the necessary condition 
of instability, 
$\partial\Omega/\partial R < 0$ (i.e. a decrease of the angular 
velocity with cylindrical radius) is fulfilled (Velikov 1959; Kurzweg 1963; Balbus 
\& Hawley 
1991; Kaisig, Tajima \& Lovelace 1992; Kumar, Coleman \& Kley 1994; 
Zhang, Diamond \& Vishniac 1994). This instability exists not only 
for short wavelength perturbations, but also for global modes with 
scales comparable to the disk height (Curry, Pudritz \& Sutherland 
1994; Curry \& Pudritz 1995). Note that the magnetic shear instability 
can arise only if the field is not too strong, because this would 
suppress the instability (Urpin 1996; Kitchatinov \& R\"{u}diger 
1997). Simulations of the magnetorotational instability in disks 
(Hawley, Gammie \& Balbus 1995; Matsumoto \& Tajima 1995; Brandenburg 
et al. 1995; Torkelsson et al. 1996; Arlt \& R\"udiger 2001) show 
that the generated turbulence may enhance the angular momentum transport. 

Unlikely, that the magnetic shear instability is the only instability
that can exist in such complex objects as astrophysical disks. A
detailed analysis of MHD modes in stratified magnetic accretion
disks demonstrates a much wider variety of instabilities than
previously realized (Keppens, Casse \& Goedbloed 2002). Therefore,
the current view on the origin of turbulence can be very simplified. 
Even a pure hydrodynamic origin of turbulence cannot be excluded
(see Richard \& Zahn 1999).  
 
The situation is particularly uncertain in cold and dense protostellar
disks where the electrical conductivity is extremely small because 
of a low ionization degree. The magnetic Reynolds number is likely not 
very large in these disks, and the magnetic field cannot be considered 
as ``frozen'' into the gas (Gammie 1996).  The behavior of the magnetic
shear instability in the presence of ohmic dissipation has been 
considered in the linear (Jin 1996) and non-linear regimes (Sano,
Inutsuka \& Miyama 1998). As it was first pointed out by Wardle 
(1999), however, poorly conducting protostellar disks can be strongly 
magnetized if electrons are the main charge carriers. Magnetization
leads to anisotropic electron transport with substantially different 
properties along and across the magnetic field (see, e.g., Spitzer
1978). If the field is sufficiently strong then the main contribution
to the electric resistivity tensor is provided by the Hall component that
 produces the electric field perpendicular to both 
the magnetic field and electric current. This component is 
non-dissipative but it can change a geometry of the magnetic field. A 
linear stability analysis undertaken by
Wardle (1999) shows that the Hall effect can provide an additional
either stabilizing or destabilizing influence depending on a direction 
of the magnetic field. A more general consideration of the
magnetic shear instability in the presence of Hall currents has been 
done by Balbus \& Terquem (2001). They found that the Hall
effect changes qualitatively the stability properties of rotating 
gas resulting in destabilization of even outwardly increasing 
differential rotation (see also R\"udiger \& Shalybkov 2003). In their 
analysis, however, Balbus \& Terquem (2001) neglected gravity which, 
in protostellar disks, is as important as rotation. Gravity influences 
the behavior of modes via 
buoyancy, and may lead to new Hall-driven instabilities
missed in simplified considerations.  
       
In the present paper we consider the linear stability properties of 
magnetic protostellar disks taking into account the Hall effect and 
gravity. We treat the behavior of different short wavelength 
magnetohydrodynamic modes which can exist in such objects and 
determine the parameter domain where these modes are unstable.
The paper is organized as follows. In Section 2, we discuss the 
Hall effect in the conditions of protostellar clouds. In Section 2, 
the main equations are presented and a dispersion relation is derived 
that describes the behavior of short wavelength perturbations in the 
Boussinesq approximation. The stability criteria for different modes 
are discussed in Section 4, and the growth rates of instabilities are
calculated in Section 5. Finally, our results are briefly summarized 
in Section 6.

\section{Anisotropic electric resistivity in protostellar disks}

\begin{figure}
\hbox{\hskip  0.5truecm
\psfig{figure=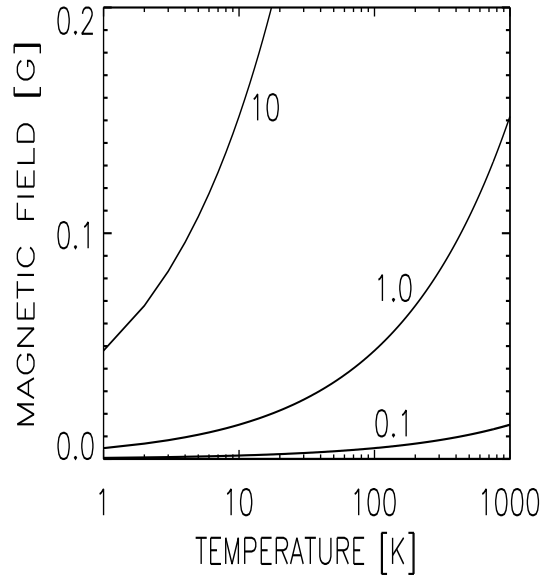,width=8cm,height=9cm}
}
\caption{The condition $a_e=1$  where the lines are marked with the 
values of the number density $n_{14}$. For magnetic fields above the 
lines the Hall effect dominates the ohmic dissipation.}
\label{plot1}
\end{figure}

The electrical conductivity is likely very low in protostellar disks 
because of a low temperature. The magnetic field cannot be considered
as ``frozen'' into gas, and dissipative effects should be taken 
into account. However, despite low temperature and ionization, the 
electron gas can be magnetized as it was first pointed out by Wardle 
(1999). Magnetization of the electron gas is characterized by the 
product of electron gyrofrequency, $\omega_{B} = e B /m_{e} c$  and the relaxation time of electrons, 
$\tau$ (see, e.g., Spitzer 1978). In protostellar disks, $\tau$ is 
likely determined by the scattering of electrons on neutrals, then 
$\tau = 1/ n \langle \sigma v \rangle$ where $\langle \sigma v 
\rangle$ is the average product of the cross-section and velocity
and $n$ is the number density of neutrals. 
Using the estimate of $\langle \sigma v \rangle$ obtained by Draine, 
Roberge \& Dalgarno (1983) for electron-neutral collisions, we can 
represent the magnetization parameter of electrons, $a_{e}$, as
\begin{equation}
a_{e} \equiv \omega_{B} \tau = 21\ B n_{14}^{-1} 
T_{2}^{-1/2},
\label{1}
\end{equation}  
where $B$ is measured in Gauss, $n_{14}$ has units $10^{14}$ 
cm$^{-3}$, and $T_{2}$ has units $100$ K. If this parameter is greater 
than 1, i.e. 
\begin{equation}
B>  0.048 \ n_{14} \
\sqrt{T_2} \;\; {\rm Gauss}
\label{1.1}
\end{equation} 
(see Fig. \ref{plot1}), then electron transport is anisotropic, and we 
have to use a tensorial magnetic diffusivity instead of 
a scalar one.  For more details concerning the generalized Ohm's law in 
weakly ionized plasma we refer to the paper by Shalybkov \& Urpin (1995) 
where this law has been considered using the relaxation time approximation 
for three components plasma with ions and neutrals of the same mass.
In very weakly ionized plasma of protostellar disks, the difference
between parallel and perpendicular resistivity is small (see, e.g.,
Balbus \& Terquem 2001). The Hall-originated  magnetic diffusivity is then given by
\begin{equation}
a_e \eta = \beta B = {cB \over 4\pi e n_{\rm e}},
\label{2}
\end{equation}  
where $\eta$ is the microscopic magnetic diffusivity and $n_{\rm e}$ is 
the number density of electrons. The magnetic diffusivity is
\begin{equation}
\eta = 2.34 \times 10^{3} f^{-1} T_{2}^{1/2} \;\; {\rm cm}^{2} \;
{\rm s}^{-1},
\end{equation} 
where $f=n_{\rm e}/n$ is an ionization fraction.

The induction equation reads
\begin{equation}
\frac{\partial \vec{B}}{\partial t} = {\rm rot} (\vec{U} \times
\vec{B}) + \eta \Delta \vec{B} - \beta {\rm rot} ({\rm rot}\vec{B} 
\times \vec{B}).
\label{3}
\end{equation}
The last term represents the Hall effect, and numerical evaluation
done by Wardle \& Ng (1999) indicate that this term can be of 
importance in some regions of protostellar disks. In
(\ref{3}), we neglect a non-uniformity of the resistivity and
electron number density. 

%%%%%%%%%%%%%%%%%%%%%%%%%%%%%%%%%%%%%%
\section{Basic equations and the dispersion relation}
%%%%%%%%%%%%%%%%%%%%%%%%%%%%%%%%%%%%%
Consider the stability properties of a magnetized axisymmetric 
protostellar disk of a finite vertical extent. The unperturbed angular 
velocity can generally depend on both $R$ and $z$, so $\Omega = 
\Omega(R, z)$, where ($R$, $\phi$, $z$) are cylindrical 
coordinates. The magnetic field, $\vec{B}=(B_{R}, B_{\phi}, 
B_{z})$, is assumed to be weak in the sense that the Alfv\'en speed, 
$c_{\rm A}$, is small compared with the sound speed, $c_{\rm s}$.
This enables us to employ the Boussinesq approximation.

In the unperturbed state, the disk is assumed to be in hydrostatic 
equilibrium in the $R$- and $z$-directions,
\begin{equation}
\frac{\nabla p}{\rho} = \vec{G} + \frac{1}{4 \pi \rho}
{\rm rot} \vec{B} \times \vec{B} \;\;, \;\;\;\;
\vec{G} = \vec{g} + \Omega^{2} \vec{R} \;,
\label{4}
\end{equation}
where $\vec{g}$ is the gravity. If $c_{\rm s} > c_{\rm A}$, the unperturbed 
Lorentz force is small compared with the pressure force, thus the 
disk structure is mainly determined by the balance between gravity,
centrifugal force, and pressure. 

We consider axisymmetric short wavelength perturbations with the 
space-time dependence $\exp(\gamma t - i \vec{k} \cdot \vec{x})$ 
where $\vec{k}= (k_{R}, 0, k_{z})$ is the wave vector, $| \vec{k} \cdot 
\vec{x} | \gg 1$. Small perturbations will be indicated by subscript 1,
whilst unperturbed quantities will have no subscript, except for 
indicating vector components when necessary. The linearized momentum 
and continuity equations governing the behavior of such perturbations 
in the Boussinesq approximation 
read
\begin{eqnarray}  
\lefteqn{\gamma \vec{U}' + 2 \vec{\Omega} \times \vec{U}' +
\vec{e}_{\phi} R (\vec{U}' \cdot \nabla) \Omega =
\frac{i \vec{k} p'}{\rho} - \alpha \vec{G} T' +}
\nonumber
\label{5} \\
&& \quad \quad \frac{i}{4\pi \rho} [\vec{k} (\vec{B} \cdot \vec{B}')
-\vec{B}' (\vec{k} \cdot \vec{B})] \;, \\
\lefteqn{\vec{k} \cdot \vec{U}' = 0 \;,}
\label{6}
\end{eqnarray}
where $\vec{U}'$, $\vec{B}'$, $p'$ and $T'$ are 
perturbations of the hydrodynamic velocity, magnetic field, pressure 
and temperature, respectively; $\alpha= - (\partial \ln \rho / 
\partial T)_{p}$ is the thermal expansion coefficient and
 $\vec{e}_{\phi}$ is the unit vector in the azimuthal direction. 
In Eq. (6) it is assumed that the density perturbation in 
the buoyancy force is determined by the temperature perturbation, 
thus $\rho'= - \rho \alpha T'$, in accordance with the idea of 
the Boussinesq approximation. 

Since the thermal conductivity of protostellar clouds is low because
of a low temperature ($T \sim 10-10^{3}$ K), we adopt the adiabatic
equation to describe the evolution of temperature perturbations,
\begin{equation}
\gamma T' + \vec{U}' \cdot (\Delta \nabla T) = 0 \;, 
\label{7}
\end{equation}
where $(\Delta \nabla T) = \nabla T - \nabla_{\rm ad} T $ is the 
difference between the actual and adiabatic temperature gradients. 

The linearized induction equation and the divergence free condition
read
\begin{eqnarray}
\lefteqn{(\gamma + \omega_{\rm R}) \vec{B}' = -i \vec{U}' (\vec{k} \cdot 
\vec{B}) +
R \vec{e}_{\phi} (\vec{B}' \cdot \nabla \Omega)  +} 
\nonumber\\
&& \quad \quad \quad \quad \quad \ + \beta (\vec{k} \cdot \vec{B}) 
\vec{k} \times 
\vec{B}',
\label{8}\\
\lefteqn{\vec{k} \cdot \vec{B}' = 0,}
\label{9}
\end{eqnarray}
where $\omega_{\rm R} = \eta k^{2}$ is the inverse timescale
of the ohmic field decay.

The dispersion equation governing the behavior of perturbations
may be obtained in the standard way. Equating the determinant of the 
set of Eqs. (\ref{5})\dots(\ref{9}) to zero, we obtain
\begin{equation}
\gamma^{5} + a_{4} \gamma^{4} + a_{3} \gamma^{3} + a_{2} \gamma^{2}
+ a_{1} \gamma + a_{0} = 0 \; , 
\label{10}
\end{equation}
where
\begin{eqnarray}
\lefteqn{a_{4} = 2 \omega_{\rm R} ,}\nonumber\\
\lefteqn{a_{3}= \omega_{\rm R}^{2} + \omega_{\rm H} ( \omega_{\rm H} + 
\omega_{\rm sh})
+ 2\omega_{\rm A}^{2} + \omega_{\rm g}^{2} + Q^{2} ,} 
\nonumber \\
\lefteqn{a_{2} = 2 \omega_{\rm R} (\omega_{\rm g}^{2} + \omega_{\rm A}^{2} + 
Q^{2}) ,}
\nonumber \\
\lefteqn{a_{1} = [\omega_{\rm R}^{2} + \omega_{\rm H} ( \omega_{\rm H} + 
\omega_{\rm sh})]
(\omega_{\rm g}^{2} + Q^{2}) +} 
\nonumber \\
&& \quad + \omega_{\rm A}^{2} \left[ \omega_{\rm g}^{2} +
\omega_{\rm A}^{2} + \omega_{\rm H} \omega_{\rm sh} + 2 \Omega 
\frac{k_{z}}{k} 
(2 \omega_{\rm H}+ \omega_{\rm sh}) \right] ,
\nonumber \\ 
\lefteqn{a_{0} = \omega_{\rm R} \omega_{\rm A}^{2} \omega_{\rm g}^{2} ,} 
\nonumber
\end{eqnarray}
and
\begin{eqnarray}
\lefteqn{Q^{2} = 4 \Omega^{2} \frac{k_{z}^{2}}{k^{2}} + 2 \Omega R
\frac{k_{z}}{k^{2}} 
\left( k_{z} \frac{\partial \Omega}{\partial R}
- k_{R} \frac{\partial \Omega}{\partial z} \right) \;,}
\nonumber \\
\lefteqn{\omega_{\rm g}^{2} = - \alpha \Delta \nabla T \cdot \left[ 
\vec{G} -
\frac{\vec{k}}{k^{2}} (\vec{k} \cdot \vec{G}) \right] \; ,}
\nonumber\\
\lefteqn{\omega_{\rm H} = \beta k (\vec{k} \cdot \vec{B}) \;,}\nonumber\\
\lefteqn{\omega_{\rm sh} = \frac{R}{k}
\left( k_{z} \frac{\partial \Omega}{\partial R}
- k_{R} \frac{\partial \Omega}{\partial z} \right) \;,}
\nonumber
\end{eqnarray}
$\omega_{\rm g}$ is the frequency of buoyancy waves; $\omega_{\rm A} =
(\vec{k} \cdot \vec{B}) / \sqrt{4 \pi \rho}$ is the Alfv\'en
frequency, $Q^{2}$ represents the effects associated with differential
rotation, $\omega_{\rm H}$ and $\omega_{\rm sh}$ are the characteristic 
frequencies of the Hall- and shear-driven processes, respectively.   

If gravity is neglected ($\omega_{\rm g}=0$) then we recover the dispersion
equation derived by Balbus \& Terquem (2002) 
\begin{equation}
\gamma^{4} + b_{3} \gamma^{3} + b_{2} \gamma^{2} + b_{1} \gamma
+ b_{0} = 0,
\label{11}
\end{equation}
where
\begin{eqnarray}
\lefteqn{b_{3} = 2 \omega_{\rm R} , \; }\nonumber\\
\lefteqn{b_{2}= \omega_{\rm R}^{2} + \omega_{\rm H} ( \omega_{\rm H} + 
\omega_{\rm sh})
+ 2\omega_{\rm A}^{2} + Q^{2} , }
\nonumber \\
\lefteqn{b_{1} = 2 \omega_{\rm R} (\omega_{\rm A}^{2} + Q^{2}) ,}
\nonumber \\
\lefteqn{b_{0} = \omega_{\rm R}^{2} Q^{2} + \left[ \omega_{\rm A}^{2} + 
2 \Omega
\frac{k_{z}}{k} (\omega_{\rm H} + \omega_{\rm sh}) \right] \times}
\nonumber \\
&& \quad \quad \quad \quad \quad \quad \left[ \omega_{\rm A}^{2}
+ \omega_{\rm H} \left( 2 \Omega \frac{k_{z}}{k} + \omega_{\rm sh} \right) 
\right]. \nonumber
\end{eqnarray}
Assuming that the condition of instability is given by 
\begin{equation}
b_{0} < 0,
\label{12}
\end{equation}
Balbus \& Terquem (2001) argued that the Hall effect can destabilize
any differential rotation laws in protostellar disks, even those with 
angular velocity increasing outward.

\section{Criteria of instability of protostellar disks}

The equation (\ref{10}) describes five low-frequency modes which can 
exist in protostellar disks. The condition that at least one of the 
roots of equation (\ref{10}) has a positive real part (unstable mode) is 
equivalent to one of the following inequalities
\begin{eqnarray}
\lefteqn{a_{0} <0 \;,}
\nonumber \\
\lefteqn{A_{1} \equiv a_{4} a_{3} - a_{2} < 0 \;, }
\nonumber \\
\lefteqn{A_{2} \equiv a_{2} (a_{4} a_{3} -a_{2}) - a_{4}(a_{4} a_{1} 
- a_{0}) <0 \;,}
\nonumber \\
\lefteqn{A_{3} \equiv (a_{4} a_{1} -a_{0}) [a_{2} (a_{4} a_{3} 
- a_{2}) - a_{4}
(a_{4} a_{1} - a_{0})] - }
\nonumber \\
&& \quad \quad \quad \quad \quad \quad - a_{0} (a_{4} a_{3} 
- a_{2})^{2} < 0 \;,
\label{13}
\end{eqnarray}
being fulfilled (see, e.g., Aleksandrov, Kolmogorov \& Laurentiev
1985). Since $\omega_{\rm R}$ is positive defined quantity, the first 
condition ($a_{4} < 0$) will never apply, and only the four given 
conditions determine the instability in disks.

\subsection{The condition $a_{0} < 0$}

Since $\omega_{\rm A}^{2} > 0$, 
the condition $a_{0} < 0$ is equivalent to 
\begin{equation}
\omega_{\rm g}^{2} < 0, 
\label{14}
\end{equation}
or
\begin{equation}
k^{2} (\vec{G} \cdot \Delta \nabla T) - (\vec{k} \cdot \vec{G})(\vec{k}
\cdot \Delta \nabla T) > 0.
\label{15}
\end{equation}
Generally, $\omega_{\rm g}^{2} < 0$ if the temperature gradient is 
superadiabatic, i.e. it exceeds 
its adiabatic value. In this case, the standard convective instability 
arises. However, $\omega_{\rm g}^{2}$ may also be negative if the 
temperature gradient is subadiabatic but $\Delta \nabla T$ is not 
parallel to the ``effective gravity", $\vec{G}$ (see Urpin \& Brandenburg 
1998). This obliqueness can be caused, in principle, either by the 
dependence of $\Omega$ on $z$ or by radiative heat transport in the 
radial direction. Introducing the angle $\psi$ between the vectors 
$\vec{G}$ and $\vec{k}$ and representing $\Delta \nabla T$ as a sum 
of components parallel and perpendicular to $\vec{G}$, $\Delta \nabla T 
= (\Delta \nabla T)_{\parallel} + (\Delta \nabla T)_{\perp}$, the 
inequality (\ref{15}) can be rewritten in the form
\begin{equation}
G (\Delta \nabla T)_{\parallel} 
[\sin^{2} \psi 
- \sin \psi \cos \psi (\Delta \nabla T)_{\perp}/ (\Delta 
\nabla T)_{\parallel}] > 0 \;. 
\label{16}
\end{equation}
If stratification is stable according to the standard Schwarzschild
criterion of convection, $G (\Delta \nabla T)_{\parallel} < 0$, then
the instability arises at
\begin{equation} 
\sin^{2} \psi 
- \sin \psi \cos \psi (\Delta \nabla T)_{\perp}/ (\Delta 
\nabla T)_{\parallel} < 0 \;. 
\label{17}
\end{equation}
This condition can be fulfilled due to the obliqueness
of $\vec{G}$ and $\Delta \nabla T$ for perturbations with a small 
(but non-zero) angle $\psi$. Estimating $(\Delta \nabla T)_{\perp}$ 
as $ (z/R) (\Delta \nabla T)_{\parallel}$, we obtain that 
$\omega_{\rm g}^{2}$ for unstable perturbations is small,
\begin{equation}
\omega_{\rm g}^{2} \sim - \Omega^{2} (H/R)^{2} \;. 
\label{18}
\end{equation}
Therefore, convection caused by obliqueness of $\vec{G}$ and $\Delta
\nabla T$ is probably relatively slow.

\subsection{The condition $A_{1} < 0$}

 This condition reads
\begin{equation}
\omega_{\rm A}^{2} + \omega_{\rm R}^{2} + \omega_{\rm H}^{2} + 
\omega_{\rm H} 
\omega_{\rm sh} < 0,
\label{19}
\end{equation}
or, substituting the frequencies,
\begin{equation}
\beta R (\vec{k}\!\!\cdot\!\!\vec{B})\!\left(\!\!k_{z}
\frac{\partial \Omega}{\partial R} - k_{R} \frac{\partial 
\Omega}{\partial z}\!\!\right)\!<  -\!\omega_{\rm A}^{2} -\!\omega_{\rm R}^{2}
- \beta^2 k^{2} (\vec{k}\!\!\cdot\!\!\vec{B})^{2}.
\label{20}
\end{equation}
Despite this inequality is the only condition (\ref{13}) that does not 
depend on gravity, it differs from the criterion of instability
derived by Balbus \& Terquem (2001) (see Eq. (85) of their paper). 
The condition (\ref{20}) describes a new instability that
appears due to combined influence of shear and the Hall effect.
This instability differs from the magnetic shear instability because 
the only term that can provide a destabilizing 
influence is proportional to the Hall frequency and shear stresses, 
and this terms is vanishing if $\omega_{\rm H} \rightarrow 0$. 
To satisfy the inequality (\ref{20}) the sign of the left hand side 
should be negative since all three terms on the right hand side are 
negative. Obviously, for any dependence of $\Omega$ on $R$ and $z$ 
and for any direction of $\vec{B}$, there exist wave vectors that  
satisfies the inequality
\begin{equation}
(\vec{k} \cdot \vec{B}) \left( k_{z}
\frac{\partial \Omega}{\partial R} - k_{R} \frac{\partial 
\Omega}{\partial z} \right) < 0,
\label{21}
\end{equation}
and makes the left hand side of Eq. (\ref{20}) negative. Therefore, 
any differential rotation can generally be unstable if the Hall 
effect is sufficiently strong. 

Compare characteristic frequencies in Eq. (\ref{19}).
If the Hall parameter is large, $a_{e} \gg 1$, then the ``ohmic 
frequency'' is negligible in Eq. (\ref{19}). The characteristic 
value of the Hall frequency, $\omega_{\rm H}$, is
\begin{equation}
\omega_{\rm H} \sim 2 \times 10^{-4} B n_{{\rm e}2}^{-1} 
\lambda_{11}^{-2} \;\;\; {\mathrm s}^{-1},
\label{22}
\end{equation}
where $\lambda=2 \pi/k$ is the wavelength, and $\lambda_{11}$ has 
units $10^{11}$ cm; $n_{{\rm e}2}$ are in units 100 cm$^{-3}$.
Assuming that gas is weakly ionized and $n \gg n_{\rm e}$, we can 
estimate
\begin{equation}
\left| \frac{\omega_{\rm H}}{\omega_{\rm A}} \right| \approx  
\frac{ck}{\omega_{\rm p}} 
\sqrt{\frac{m_{\rm p} n}{m_{\rm e} n_{\rm e}}} 
\approx 1.4 \times 10^{-4} \lambda_{11}^{-1} n_{\rm e2}^{-1/2} f^{-1/2}
%43 \left( \frac{ck}{\omega_{p}} \right) f^{-1/2},  
\label{23}
\end{equation}
where $\omega_{\rm p} = ( 4 \pi e^{2} n_{\rm e}/m_{\rm e})^{1/2}$ is 
the plasma 
frequency, $m_{\rm e}$ and $m_{\rm p}$ are the electron and proton mass, 
respectively. Since $f$ can be very small in a large fraction of the 
disk volume ($f \sim 10^{-(11...13)}$), most likely that $|\omega_{\rm H}| 
> |\omega_{\rm A}|$. In this 
case, the Alfv\'en frequency is small in Eq. (\ref{19}). Then, in a row 
with Eq. (\ref{21}), the necessary condition of instability is also
\begin{equation}
|\omega_{\rm sh}| > |\omega_{\rm H}|.
\label{24}
\end{equation} 
Assuming $|\partial \Omega / \partial R| \sim \Omega/R$ as is usual 
in astrophysical disks, we can represent the necessary condition 
(\ref{24}) as
\begin{equation}
B < {\Omega \over \beta k^2} \simeq 10^{-3} P_{\rm yr}^{-1} n_{\rm e2} 
\lambda_{11}^{2} \;\; {\rm Gauss},
\label{25}
\end{equation}
where $P= 2 \pi /\Omega$ is the rotation period, $P_{\rm yr}$ has units
1 yr. The condition that the Hall effect dominates
ohmic dissipation, $a_{e} >1$, yields
\begin{equation}
B > 0.048\ n_{14}\ T_{2}^{1/2} \;\; {\rm Gauss}.
\label{26}
\end{equation}
The conditions (\ref{25}) and (\ref{26}) are consistent only if
\begin{equation}
P_{\rm yr} < 2.2 \times 10^{-2} f_{-12} \lambda_{11}^{2} 
T_{2}^{-1/2},
\label{27}
\end{equation} 
where $f_{-12}= f/10^{-12}$. If we estimate the thickness of a disk,
$H$, at the distance $R$ as $H \sim 0.1 R$ and assume that $\lambda 
\sim 0.1 H$ to justify a short wavelength approximation, then we have 
at $R=1$ AU,   
\begin{equation}
P_{\rm yr} < 2.2 \times 10^{-2} f_{-12} T_{2}^{-1/2}.
\label{28}
\end{equation} 
To be fulfilled this condition requires the temperature lower than
100 K and ionization higher than $10^{-12}$. Therefore, the condition 
(\ref{27}) seems to be very restricting and hardly to be fulfilled in the 
conditions of protostellar disks.

\subsection{The condition $A_{2} < 0$}

Substituting the expressions for
coefficients, we can represent the condition $A_{2} < 0$ as
\begin{equation}
\omega_{\rm g}^{2} + 2 \omega_{\rm R}^{2} + 
2 \left( 2 \Omega \; \frac{k_{z}}{k} - \omega_{\rm H} \right)^{2} < 0.
\label{29}
\end{equation}
All terms on the left hand side are positive except $\omega_{\rm g}^{2}$ 
and, hence, the criterion (\ref{29}) can be satisfied only if 
$\omega_{\rm g}^{2} 
< 0$. However, this criterion requires larger negative $\omega_{\rm g}^2$
than the condition (\ref{14}). Therefore, the criterion $A_{2} < 0$
can be fulfilled only if the condition $a_{0} < 0$ is already
fulfilled.

\subsection{The condition $A_{3} < 0$}

 This condition generalizes
the criterion obtained by Balbus and Terquem (2001) for the
protostellar disk with no gravity. If we assume $\vec{g} =0$ 
(and, hence, $\omega_{\rm g}^{2} = a_{0}=0$) then $A_{3} = a_{4} 
a_{1} A_{2}$. Since $a_{4}$ is always positive and $A_{2}$ is 
positive at $\vec{g}=0$ (see Eq. (\ref{29})), the criterion $A_{3} 
< 0$ reduces to
\begin{equation}
\left. a_{1} \right|_{g=0} \equiv b_{0} <0,
\label{30}
\end{equation}
that is the condition derived by Balbus \& Terquem (2001).

In the general case, the criterion $A_{3} < 0$ yields
\begin{eqnarray}
D \equiv b_{0} + q \left\{ (\omega_{\rm R}^{2} + \omega_{\rm H}^{2} + 
\omega_{\rm H} \omega_{\rm sh}) \left[ \omega_{\rm g}^{2} + Q^{2} - 
\omega_{\rm A}^{2}
\right. \right.
\nonumber \\
\left. \left.
+ \omega_{\rm R}^{2} + 2 \left( 2 \Omega
\frac{k_{z}}{k} - \omega_{\rm H} \right)^{2}  
- \omega_{\rm H}^{2} - \omega_{\rm H} \omega_{\rm sh}
\right] \right.
\nonumber \\
\left. + \omega_{\rm A}^{2} \left( \frac{1}{2} \omega_{\rm g}^{2} + Q^{2}
\right) \right\} < 0,
\label{31}
\end{eqnarray}
where
\begin{equation}
q = \frac{1}{2} {\omega_{\rm g}^{2} \over \omega_{\rm R}^{2} + 
\left( 2 \Omega
\frac{k_{z}}{k} -\omega_{\rm H} \right)^{2}}.
\label{32}
\end{equation}
Gravity can provide either positive or negative contribution to the 
left hand side of Eq. (\ref{31}) and, hence, can be either a 
stabilizing or destabilizing factor depending on the characteristic 
frequencies. The condition (\ref{31}) is generally rather cumbersome, and 
we consider only the particular cases of astrophysical interest. 

In protostellar disks, we have typically $\omega_{\rm g} \sim \Omega$ 
and, hence, $q \sim 1$. As it has been adopted in previous studies,
we also assume that the Hall effect dominates ohmic dissipation, and 
$\omega_{\rm H} > \omega_{\rm R}$ (or $a_{e} >1$),
\begin{equation}
\omega_{\rm R} \approx 9.2 \times 10^{-6} f_{-12}^{-1} T_{2}^{1/2}
\lambda_{11}^{-2} \;\; {\rm s}^{-1}.
\end{equation}
Stability properties are sensitive to the relationship between the Hall 
frequency, $\omega_{\rm H}$, and the angular velocity, $\Omega$. The 
inequality $\omega_{\rm H} > \Omega$ is fulfilled only if the field is 
strong enough, 
\begin{equation}
B \gg 10^{-3} n_{{\rm e}2} \lambda_{11}^{2} P_{\rm yr}^{-1}.
\label{33}
\end{equation}
This field seems to be rather strong for the conditions of protostellar
disks and, most likely, $\Omega \gg \omega_{\rm H}$. Note also that
at $\omega_{\rm H} > \Omega$ we have $D > 0$ for both $\omega_{\rm H} >
\omega_{\rm A}$ and $\omega_{\rm A} > \omega_{\rm H}$. Hence, the instability
does not appear, and the field, satisfying Eq. (\ref{33}), is 
stabilizing for instability given by the criterion $A_{3} < 0$.
Therefore, we consider Eq. (\ref{31}) only at $\Omega \gg \omega_{\rm H}$.

If this is the case, the criterion of instability reads
\begin{eqnarray}
\omega_{\rm H} (\omega_{\rm H} + \omega_{\rm sh}) \left[ (1+q) 
( \omega_{\rm g}^{2} 
+ Q^{2}) - q \omega_{\rm A}^{2} \right] + \omega_{\rm A}^{4}+
\nonumber \\
+ \omega_{\rm A}^{2} \left[ \omega_{\rm H} \left(
2 \Omega \frac{k_{z}}{k} + \omega_{\rm sh} \right) 
+ 2 \Omega \frac{k_{z}}{k} ( \omega_{\rm H} + \omega_{\rm sh} ) \right]+
\nonumber \\
+ q \omega_{\rm A}^{2}
\left( \frac{1}{2} \omega_{\rm g}^{2} + Q^{2} \right) < 0. 
\label{34}
\end{eqnarray}
As it was mentioned (see Eq. (\ref{23})), the Hall frequency can 
exceed the Alfv\'en frequency in protostellar disks, $\omega_{\rm H} > 
\omega_{\rm A}$, and we treat this case first. Under this 
assumption, the criterion (\ref{34}) yields  
\begin{equation}
\omega_{\rm H} ( \omega_{\rm H} + \omega_{\rm sh} ) (1+q) 
( \omega_{\rm g}^{2} 
+ Q^{2}) < 0.
\label{35}
\end{equation}
In a convectively stable disk with $\omega_{\rm g}^{2} > 0$ and $Q^{2}>0$, 
the shear-driven instability arises only if 
\begin{equation}
\omega_{\rm H} ( \omega_{\rm H} + \omega_{\rm sh} ) < 0.
\label{36}
\end{equation} 
To be fulfilled, this inequality requires a strong 
shear, satisfying Eq. (\ref{24}), and a wave vector directed in 
accordance with Eq. (\ref{21}). There is no instability due to 
the Hall effect without shear.
 It appears that the condition 
$\omega_{\rm sh} > \omega_{\rm H}$ is general for instabilities caused by 
a joint influence of the Hall effect and shear but, as we already 
mentioned, it is very restricting in protostellar disks. The sign 
of $\omega_{\rm sh}$ plays no crucial role in the condition (\ref{36}), and 
instability can appear for any $\partial \Omega / \partial R$ if 
Eq. (\ref{24}) is fulfilled. Note that in the case $\vec{g}=0$, the 
instability can arise for any sign of shear as well (Balbus \&
Terquem 2001).  

Consider now the criterion of instability (\ref{34}) in the case when the 
Hall frequency is small compared to the Alfv\'en frequency, $\omega_{\rm A} 
> \omega_{\rm H}$. We again assume $\Omega > \omega_{\rm H}$ but the Alfv\'en
frequency can generally be larger or smaller than $\Omega$. If
$\omega_{\rm A} \gg \Omega$, or
\begin{equation}
B > 0.15\ n_{14}^{1/2}\ \lambda_{11}\ P_{\rm yr}^{-1},
\label{37}
\end{equation}  
then $D \approx \omega_{\rm A}^{4} > 0$, and instability is not 
possible. The field, satisfying this condition, is typically 
even stronger than that given by  (\ref{33}). Unlikely that
such a strong field can exist in protostellar disks, and we 
consider instability under a more realistic condition $\Omega \gg 
\omega_{\rm A}$. 

If $\Omega > \omega_{\rm A} > \Omega \sqrt{\omega_{\rm H}/ \Omega}$, then 
we have from the condition (\ref{34}) 
\begin{equation}
(1+q) \left( \omega_{\rm g}^{2} + 4 \Omega \frac{k_{z}}{k} \omega_{\rm sh}
\right) < 0.
\label{38}
\end{equation}
In a convectively stable disk, the instability appears if
\begin{equation}
4 \Omega \frac{k_{z}}{k} \omega_{\rm sh} + \omega_{\rm g}^{2} < 0. 
\label{39}
\end{equation}
At $\omega_{\rm g} \sim \Omega \gg \omega_{\rm A}$, this condition is far 
more restricting than the criterion of the magnetorotational 
instability obtained by Balbus \& Terquem (2001) for protostellar
disks.

If $\Omega \sqrt{\omega_{\rm H}/ \Omega} > \omega_{\rm A}$, then we 
recover the
instability criterion (\ref{35}).

\section{The growth rate of instability}

A general analysis of the roots of Eq. (\ref{10}) is very complicated. 
However, simple expressions for the roots can be obtained in some
cases of astrophysical interest. Consider initially Eq. (\ref{10}) in 
the case when the ohmic dissipative ``frequency'', $\omega_{\rm R}$, 
is small compared to other characteristic frequencies. Then, with 
accuracy in terms of the zeroth order in $\omega_{\rm R}$, four roots 
corresponding to rapidly varying modes (either growing, or decaying, 
or oscillating) can be obtained from Eq. (\ref{10}) with neglected 
ohmic dissipation. These roots satisfy the approximate equation
\begin{equation}
\gamma^{4} + c_{2} \gamma^{2} + c_{0} = 0, 
\label{40}
\end{equation} 
where
\begin{equation}
c_{2} \approx \omega_{\rm g}^{2} + Q^{2}
\label{40.1}
\end{equation}  
\begin{equation}
c_{0}= \omega_{\rm A}^{2} \left( \omega_{\rm g}^{2} + 2 \Omega 
\frac{k_{z}}{k} \omega_{\rm sh} \right)
+ \omega_{\rm H} (\omega_{\rm H} + \omega_{\rm sh}) (\omega_{\rm g}^{2} 
+ Q^{2}).
\label{41}
\end{equation}
The fifth mode describes a secular instability and varies on a long
time scale proportional to $\omega_{\rm R}$. For this mode, we have with 
the accuracy in terms of the lowest order in $\omega_{\rm R}$,
\begin{equation}
\gamma_{5} \approx - \frac{a_{0}}{a_{1}} \approx - \frac{1}{c_{0}}
\omega_{\rm R} \omega_{\rm A}^{2} \omega_{\rm g}^{2}. 
\label{42}
\end{equation}
If $\omega_{\rm A}^{2} \gg | \omega_{\rm H} (\omega_{\rm H} +
\omega_{\rm sh})|$, then Eq. (46) yields a simple estimate for $\gamma_{5}$,
\begin{equation}
\left|\frac{\gamma_{5}}{\Omega} \right| \sim 44 \; f_{-12}^{-1}
T_{2}^{1/2} \lambda_{11}^{-2} P_{\rm yr}.
\end{equation}
 
The solution of Eq. (\ref{40}) is given by
\begin{equation}
\gamma^{2} = - \frac{c_{2}}{2} \pm \sqrt{ \frac{c_{2}^{2}}{4}
- c_{0}}.
\label{43}
\end{equation}
Note that the true expressions for $\gamma$ have to contain also small 
corrections, $\Delta \gamma$, proportional to $\omega_{\rm R}$ which are
neglected in Eq. (48).
If $\Omega > \max(\omega_{\rm H}, \omega_{\rm A})$ as the criteria of 
instability requires then $c_{2}^{2} \gg 4 c_{0}$, and we 
obtain 
\begin{equation}
\gamma_{1,2}^{2} \approx - \frac{c_{0}}{c_{2}} , \quad \quad
\gamma_{3,4}^{2} \approx - c_{2}. 
\label{44}
\end{equation} 
The modes 1 and 2 are relevant to the Hall-driven instability
whereas the modes 3 and 4 describe convection and are stable in
convectively stable disks. One of the modes $\gamma_{1,2}$ is
unstable if $c_{0} < 0$. Note that the secular mode (\ref{42}) is unstable
as well under this condition, and there exist two qualitatively 
different instabilities represented by the same criterion.

In the case $\Omega \sqrt{\omega_{\rm H}/ \Omega } > \omega_{\rm A}$, we 
have for the Hall-driven modes 
\begin{equation}
\gamma_{1, 2}^{2} \approx - \omega_{\rm H} (\omega_{\rm H} + 
\omega_{\rm sh}).
\label{45}
\end{equation}
If the condition (\ref{36}) is fulfilled then one of these modes is
unstable. Since $\omega_{\rm sh} \sim \Omega$, we can estimate
\begin{equation}
\left| \frac{\gamma_{1,2}}{\Omega} \right| \sim 
\sqrt{\frac{\omega_{\rm H}}{\Omega}} \sim 31 \; B^{1/2} n_{{\rm e}2}^{-1/2}
\lambda_{11}^{-1} P_{\rm yr}^{1/2}. 
\end{equation}

If $\Omega > \omega_{\rm A} > \Omega \sqrt{ \omega_{\rm H} /\Omega}$ then 
we obtain
\begin{equation}
\gamma_{1,2}^{2} \approx - \frac{\omega_{\rm A}^{2}}{\omega_{\rm g}^{2} 
+ Q^{2}}
\left(\omega_{\rm g}^{2} + 2 \Omega \frac{k_{z}}{k} \omega_{\rm sh} \right).
\label{46}
\end{equation}
This is the known dispersion equation for  magnetic shear-driven 
modes (see, e.g., Urpin 1996) in the limit $\omega_{\rm g} \gg 
\omega_{\rm A}$, and one of these modes is unstable if
\begin{equation} 
2 \Omega \frac{k_{z}}{k} \omega_{\rm sh} < - \omega_{\rm g}^{2}.
\label{47}
\end{equation}
The growth rate is of the order of $\omega_{\rm A}$. Note that the
condition (53) differs from the more general condition (\ref{39}) 
that predicts instability also within the range
\begin{equation}
- \omega_{\rm g}^{2} < 2 \Omega \frac{k_{z}}{k} \omega_{\rm sh}  < - 
\frac{1}{2} 
\omega_{\rm g}^{2}.
\label{48}
\end{equation}
The difference originates from the approximate expressions (\ref{44}) 
that describes $\gamma_{1,2}^{2}$ neglecting terms proportional 
$\omega_{\rm R}$. If the inequality (\ref{47}) is not fulfilled, we have 
$c_{0} > 0$, and the roots $\gamma_{1}$ and $\gamma_{2}$ become 
imaginary. Therefore, their stability at $c_{0} > 0$ is determined 
by small real corrections $\propto \omega_{\rm R}$. These corrections 
can be calculated by making use of the standard perturbation 
procedure. For the modes 1 and 2, corrections are same and given 
by
\begin{equation}
\Delta \gamma_{1,2} \approx \frac{1}{2 c_{0}} (a_{0} - 2 a_{4} c_{0})
= - \frac{\omega_{\rm R} \omega_{\rm A}^{2}}{2 c_{0}} \left( 4 \Omega 
\frac{k_{z}}{k} \omega_{\rm sh} + \omega_{\rm g}^{2} \right).
\label{49}
\end{equation} 
These corrections are positive under the condition (\ref{48}) and leads 
to instability in complete agreement with the criterion (\ref{39}). 
However, the instability caused by small resistive
corrections is oscillatory and qualitatively different from the
magnetic shear instability. Note that within the range (\ref{48}) the
secular mode (\ref{42}) is stable.  

Consider now the roots of Eq. (\ref{10}) assuming that the 
dissipative ``frequency'', $\omega_{\rm R}$, is large compared to other 
characteristic frequencies including $\omega_{\rm H}$. Most likely,
that the angular velocity, $\Omega$, is largest among other
frequencies, and the assumption $\omega_{\rm R} > \Omega$ implies
that
\begin{equation}
P_{\rm yr} > 2.2 \times 10^{-2} f_{-12} \lambda_{11}^{2}  
T_{2}^{-1/2}.
\label{50}
\end{equation}
Probably, such inequality can be fulfilled in some regions of
protostellar disks if ionization is low. If $\Omega > \omega_{\rm A}$, 
then the coefficients of Eq. (\ref{10}) are given by
\begin{eqnarray}
\lefteqn{a_{4} \approx 2 \omega_{\rm R} , \;\; a_{3} \approx 
\omega_{\rm R}^{2} , \;\; 
a_{2} \approx a_{4}(\omega_{\rm g}^{2} + Q^{2}),} \nonumber\\
\lefteqn{a_{1} \approx 
\omega_{\rm R} a_{2}/2, \ \ \ 
a_{0} \approx \omega_{\rm R} \omega_{\rm A}^{2} \omega_{\rm g}^{2}.} 
\nonumber   
\end{eqnarray} 
Four roots of the Eq. (\ref{10}) with such coefficients have negative
real parts, and only one root can correspond to instability. With the 
accuracy in terms of the lowest order in $\omega_{\rm R}^{-1}$, this root 
is given by
\begin{equation}
\gamma \approx - \frac{\omega_{\rm A}^{2}}{\omega_{\rm R}} \cdot 
\frac{\omega_{\rm g}^{2}}{\omega_{\rm g}^{2} + Q^{2}}.
\label{51}
\end{equation} 
The growth rate is positive if $\omega_{\rm g}^{2} < 0$. As it was 
argued above (see Eq. (\ref{14})), the instability can appear under 
this condition if $\vec{G}$ and $\Delta \nabla T$ are not parallel. 
In protostellar disks, the growth rate of this instability is 
probably relative small, $\gamma \sim \omega_{\rm A} (\omega_{\rm A}/
\omega_{\rm R}) (H/R)^{2}$. This instability, however, can exist if 
ionization is extremely low, and the magnetic Reynolds number is 
small even for rotation. 

\section{Conclusions}

We have considered the stability properties of protostellar disks.
Compared to other studies we took into account consistently gravity
and buoyancy forces which play an important role in dynamics of disks.
It has been shown that, under certain conditions, magnetic protostellar
disks are unstable to different kinds of shear-driven 
instabilities. Some of these instabilities are relevant to the
Hall effect but some can manifest themselves even if the Hall
currents are negligible. All considered instabilities are influenced
by the magnetic field, and a relatively strong field can suppress 
instabilities. The strength of the field which can stabilize 
the flow depends generally on the conditions in disks and on the 
wavelength of the perturbation. The stabilizing field is determined
by the conditions (\ref{33}) and (\ref{37}) and is typically 
relatively strong.

The type of instability that is more efficient depends very much on
the conditions in protostellar disks and the wavelength of perturbations,
and can be different in different regions of the disk. The instability
associated with the criterion (\ref{14}) seems to be most general in
protostellar disks. It can occur both in magnetized ($a_{e} > 1$) and
non-magnetized ($a_{e} < 1$) regions and requires only non-parallel 
$\vec{G}$ and $\Delta \nabla T$. The growth rate of this instability is 
given by equations (46) and (57) in the limiting cases of small and 
large $\omega_{R}$, respectively. Note that the baroclinic instability 
appears if dissipative processes are taken into account and, therefore, 
it can also exist in accretion disks. In the present study, the baroclinic 
convection is caused by the electrical resistivity and magnetic field but 
it can also exist, for example, in non-magnetic disks if one takes account 
of the thermal conductivity (see Urpin \& Brandenburg 1998). The 
baroclinic convection, however, can be not the most efficient instability   
operating in protostellar disks since the inequality (\ref{14}) is 
fulfilled only for a very narrow range of $\vec{k}$ almost parallel to 
$\vec{G}$. 
\begin{figure}
\hbox{\hskip  0.5truecm
\psfig{figure=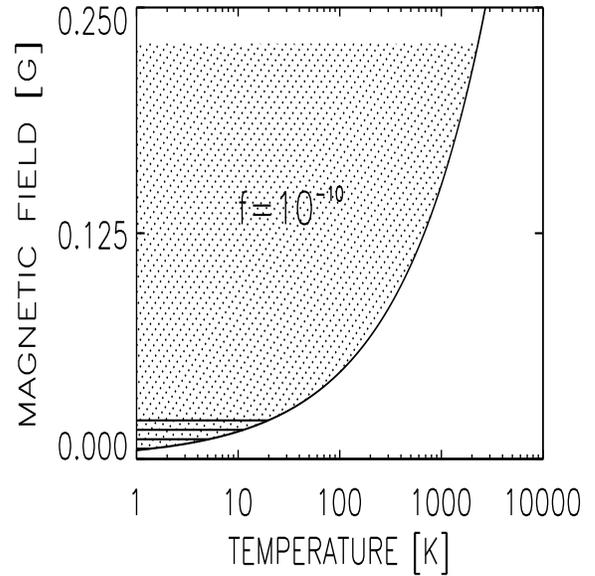,width=8cm,height=9cm}
}
\caption{The same as in Fig. \ref{plot1} but only for $n_{14}=1$......}
\label{plot2}
\end{figure}
In magnetized regions with $a_{e} > 1$, other instabilities can manifest 
themselves as well if the wavelength of perturbations is sufficiently long.
For the purpose of illustration,we consider the domains of different
instabilities in the case when the wave vector is not parallel or perpendicular 
to $\vec{\Omega}$ and $\vec{B}$ and, hence, $k_{z} \sim k$ and $(\vec{k} \cdot 
\vec{B}) \sim kB$. Then, the regions of instability are determined by two 
characteristic wavelength, $\lambda_{\rm H}$ and $\lambda_{\rm A}$, 
corresponding to the conditions $\omega_{\rm H} = \Omega$ and $\omega_{\rm A}= 
\Omega$, respectively. These wavelengths are
\begin{equation}
\lambda_{\rm H} = 3.1 \times 10^{12} B^{1/2} n_{{\rm e}2}^{-1/2}
P_{\rm yr}^{1/2} \;\; {\rm cm},
\end{equation}
\begin{equation}
\lambda_{\rm A} = 6.6 \times 10^{11} B n_{14}^{-1/2} P_{\rm yr}
\;\; {\rm cm},
\end{equation}
and their ratio is given by
\begin{equation}
\xi = \frac{\lambda_{\rm A}}{\lambda_{\rm H}} \approx 0.21 B^{1/2}
f_{-12}^{1/2} P_{\rm yr}^{1/2}. 
\end{equation}
If $\xi \gg 1$ then for $\lambda \gg \lambda_{\rm A}$ (that is equivalent to
$\Omega \gg \omega_{\rm A} \gg \Omega \sqrt{\omega_{\rm H}/\Omega}$) a
particular case of the magnetic shear instability may occur with the
growth rate given by Eq. (52). If $\xi \ll 1$ and $\lambda \gg 
\lambda_{\rm H}$ (that is equivalent to $\Omega \gg \Omega 
\sqrt{\omega_{\rm H}/\Omega} \gg \omega_{\rm A}$) then the Hall-driven 
shear instability can arise with the growth rate (50). Note that the 
both these instabilities co-exist with the baroclinic convection in 
their domains of instability but the latter can also occur for
shorter $\lambda$. 

In the present paper, we have addressed the behavior of only 
axisymmetric perturbations. It is clear, however, that the obtained
results can apply to nonaxisymmetric perturbations with the azimuthal
wavelength much longer than the vertical or radial ones, $\min (k_{R},
k_{z}) \gg k_{\phi}$. The turbulence that could be generated by 
the considered instabilities may be strongly anisotropic in the
$(R,z)$-plane, because the instability criteria are very sensitive
to the direction of the wave vector. However, the generated turbulence
may be efficient in the radial transport of angular momentum.

\section*{Acknowledgments}
The financial support by the Deutsche Forschungsgemeinschaft 
(436 RUS 113/559) is cordially acknowledged. One of the authors
(V.U.) also acknowledges financial support by the Russian
Foundation of Basic Research (grant 00-02-04011).

{}

\end{document}